\documentclass[12pt,a4paper]{article}
\usepackage{amsfonts,latexsym}
\usepackage{amsmath,amssymb}
\usepackage{graphicx,color}

\renewcommand{\thefootnote}{\fnsymbol{footnote}}

\begin{document}

\vspace{12mm}

\begin{center}
{{{\Large {\bf Shadow radius and classical scattering analysis  of two secondary hair Horndeski black holes }}}}\\[10mm]

{Yun Soo Myung\footnote{e-mail address: ysmyung@inje.ac.kr}}\\[8mm]

{Center for Quantum Spacetime, Sogang University, Seoul 04107, Republic of  Korea\\[0pt] }

\end{center}
\vspace{2mm}

\begin{abstract}
We perform the shadow radius analysis of a charged Horndeski black hole (CHB) and the naked singularity (NS) with secondary scalar hair obtained from the Einstein-Horndeski-Maxwell theory.
For this analysis, we include the beyond Horndeski black hole (bH) with secondary scalar hair  and the magnetically charged black hole (MC) found from the Einstein-Euler-Heisenberg theory.
 It is worth noting that  the NS versions of  CHB and bH arise from the charge extension of their photon spheres, while there is no NS version for MC.
 One branch (i) from the CHB is a point in the horizon realization but it shows up on the photon sphere and shadow radius.
 The shadow radius  for the CHB  is the nearly same as that for the MC with a single horizon and the charge of the NS is constrained by the EHT obserbation.
 From classical scattering analysis, it turns out that i-NS  and NS play  different roles  from  CHB, bH, and MC.
\end{abstract}
\vspace{5mm}

\vspace{1.5cm}

\hspace{11.5cm}{Typeset Using \LaTeX}
\newpage
\renewcommand{\thefootnote}{\arabic{footnote}}
\setcounter{footnote}{0}

%%%% Introduction %%%%

\section{Introduction}

It is strongly suggested  that supermassive black holes founded at the center of galaxies have played the important role in galaxy formation and galaxy evolution.
The images of the M87* BH~\cite{EventHorizonTelescope:2019dse,EventHorizonTelescope:2019ths,EventHorizonTelescope:2019ggy} have inspired enormous  studies on the  BH.
The recent EHT observation has concentrated  on  the center of our galaxy and delivered   promising  images of the  SgrA* BH~\cite{EventHorizonTelescope:2022wkp,EventHorizonTelescope:2022wok,EventHorizonTelescope:2022xqj}.
The BH images showed  that  a dark central region is  surrounded by a bright ring called shadow cast  and photon ring of the BH, respectively.
The size of the shadow corresponds to the photon sphere size additionally increased by bending of light rays, amounting to the size of the photon ring.
The shadow of BH with scalar hair was used  to test the EHT results~\cite{Khodadi:2020jij}, while the shadows of other BHs, wormholes, and naked singularities  obtained  from  modified gravity theories have been selected  to constrain their hair parameters~\cite{Vagnozzi:2022moj}.

Horndeski gravity~\cite{Horndeski:1974wa} was  regarded  as the most general scalar-tensor theory of  gravitation in four dimensional spacetime, yielding second-order field equations without ghosts.
Among  many kinds of Horndeski gravity, an important thing  is to include the nonminimal derivative coupling between scalar and Einstein tensor.
Various  black hole solutions  were found  from this gravity ~\cite{Rinaldi:2012vy,Babichev:2013cya,Anabalon:2013oea,Maselli:2015yva}.
An interesting black hole obtained from the Einstein-Horndeski-Maxwell (EHM) theory is the charged Horndeski black hole (CHB) with electric charge $q\in$[0,1.06]~\cite{Cisterna:2014nua,Feng:2015wvb}, which  implies  the presence of the secondary scalar hair $\phi(r)$ and
the existence of the naked singularity (NS) for $q\in(1.06,\infty)$~\cite{Myung:2025wmw}.

 At this stage, we wish to distinguish between primary and secondary scalar hairs.
A primary scalar hair  contains an independent scalar charge, but a secondary scalar hair does not involve  any independent scalar charge.
Furthermore, recent achievements have allowed for the beyond Horndeski  gravities~\cite{Gleyzes:2014dya,Gleyzes:2014qga,Crisostomi:2016tcp,Babichev:2017guv,Kobayashi:2019hrl}.
The BH solutions with primary scalar hair have derived from the shift and parity-symmetric subclass of beyond Horndeski gravities~\cite{Bakopoulos:2023fmv,Baake:2023zsq,Bakopoulos:2023sdm}.
On later, the regular (Bardeen) BH  solution was found from the this theory~\cite{Bakopoulos:2024ogt}. We wish to call it the beyond Horndeski  black hole (bH) with secondary scalar hair.

In this work, we wish to  perform the shadow radius analysis of  CHB, NS, and bH.
The shadow radius analysis for CHB   was recently done in~\cite{Gao:2023mjb}, but this analysis  is incomplete in the sense that it includes the single horizon $r_+(M=1,Q)$ and  single photon sphere radius $r_{ph}(1,Q)$ for $Q<1$.
Here, the electric charge  $q\in[0,1.06)$-range is allowed for the CHB, the i-NS appears at $q=1.06$, and  the $q\in(1.06,\infty)$-range is allocated for the NS  by considering  three horizons $r_1(m=1,q),r_2(1,q), r_i(1,q)$ and three photon sphere radii $L_1(1,q),L_2(1,q),L_i(1,q)$.

For a complete analysis, we with to include  a magnetically charged black hole (MC) obtained from the Einstein-Euler-Heisenberg theory~\cite{Yajima:2000kw} because it is very similar to the CHB  when its coupling constant $\mu$ is chosen to be  0.3 for shadow radius analysis~\cite{Allahyari:2019jqz,Vagnozzi:2022moj,daSilva:2023jxa}. The reason why one chooses $\mu=0.3$ is three-fold: (i)  Most of the effect on the shadow radius came from the magnetic charge rather than $\mu$. (ii) $\mu\sim 0.3$ is approximately the largest allowed coupling before the perturbative approach to the theory around the Maxwell term for $\mu\to0 $ ceases to be meaningful. (iii) If $\mu\le 0.08$, there exist four horizons~\cite{Myung:2025zxu}.
For the MC with $\mu=0.3$, there is no limitation on the magnetic charge $q$ for its horizon size, photon sphere radius, and shadow radius.

We note that two NS versions of CHB and bH  arise from the charge extension of their photon sphere except the NS and MC.
The i-NS branch from CHB  is a point in the horizon realization but it shows up on the photon and shadow radius as its NS version.
The shadow radius for the CHB  is the nearly same as that for the MC in the charge range $q\in[0,1.01]$.
Interestingly, the charge $q$ of the NS obtained from the EHM theory is  constrained when comparing the resent EHT observation.
From classical (geometric) scattering analysis, it is found  that CHB, MC, and  bH are quite different from  i-NS branch and NS.

The remainder of the present work is organized as follows.
In Sec. 2, we introduce the singular CHB, the NS, the i-NS, the regular bH, and the singular  MC  without limitation on the charge  $q$.
Sec. 3 is focussed   on computing  the shadow radii of these, depending on the charge $q$.   Two NS versions of  CHB and bH arise from the charge extension of their photon spheres, but there is no NS version for MC.
We test the seven cases including two NS versions  with the recent EHT observation in Sec. 4.
In Sec. 5, we introduce the classical scattering analysis to distinguish the CHB, bH, and MC from the i-NS  and NS. The  geometric cross sections for the i-NS  and NS blow-up at $q=1.06$ and they are divergent at $m=1.01$ ($m$: mass), showing the feature of singularities.
Finally, we discuss our results in Sec.6.

\section{Two Horndeski theories with black holes}

Firstly, we wish to consider  the Einstein-Horndeski-Maxwell (EHM)  theory with $G=1$ and a coupling constant $\gamma$~\cite{Cisterna:2014nua,Feng:2015wvb,Hajian:2020dcq}
\begin{equation}\label{Horndeski-L}
\mathcal{L}_{\rm EHM}=\frac{1}{16 \pi}\Big[R-\mathcal{F}+2\gamma G^{\mu\nu}\partial_\mu\phi\partial_\nu\phi\Big],
\end{equation}
which  respects both shift and parity symmetry of a scalar field $\phi$.
Here, $G_{\mu\nu}$ denotes the Einstein tensor and  $\mathcal{F}=F_{\mu\nu}F^{\mu\nu}$ is the Maxwell invariant term.
Considering  $G^{\mu\nu}\partial_\mu\phi\partial_\nu\phi=XR+(\square \phi)^2-\phi_{;\mu\nu} \phi^{;\mu\nu}$  up to a total divergence with $X=-(\partial \phi)^2/2$, one finds that the Einstein equation is given by $G_{\mu\nu}=\gamma T_{\mu\nu}^\phi+T_{\mu\nu}^M$ and the scalar equation takes the from $\nabla_\mu(\gamma G^{\mu\nu}\partial_\nu \phi)=0$.
This theory shows  an asymptotically flat charged Horndeski black hole (CHB) solution,
\begin{equation}\label{metric-ansatz}
ds^2_{\rm CBH}=-g(r)dt^2+\frac{dr^2}{f(r)}+r^2(d\theta^2+\sin^2\theta d\varphi^2),
\end{equation}
where two metric functions $g(r)$ and $f(r)$ are given by
\begin{align} \label{g-metricf}
g(r)=1-\frac{2m}{r}+\frac{q^2}{r^2}-\frac{q^4}{12r^4}\equiv 1-\frac{2m_f(r)}{r}, \quad f(r)=\frac{4r^4g(r)}{(2r^2-q^2)^2}
\end{align}
with the mass function
\begin{equation}
m_f(r)=m-\frac{q^2}{2r}+\frac{q^4}{24r^3}.
\end{equation}
\begin{figure*}[t!]
   \centering
   \includegraphics[width=0.5\textwidth]{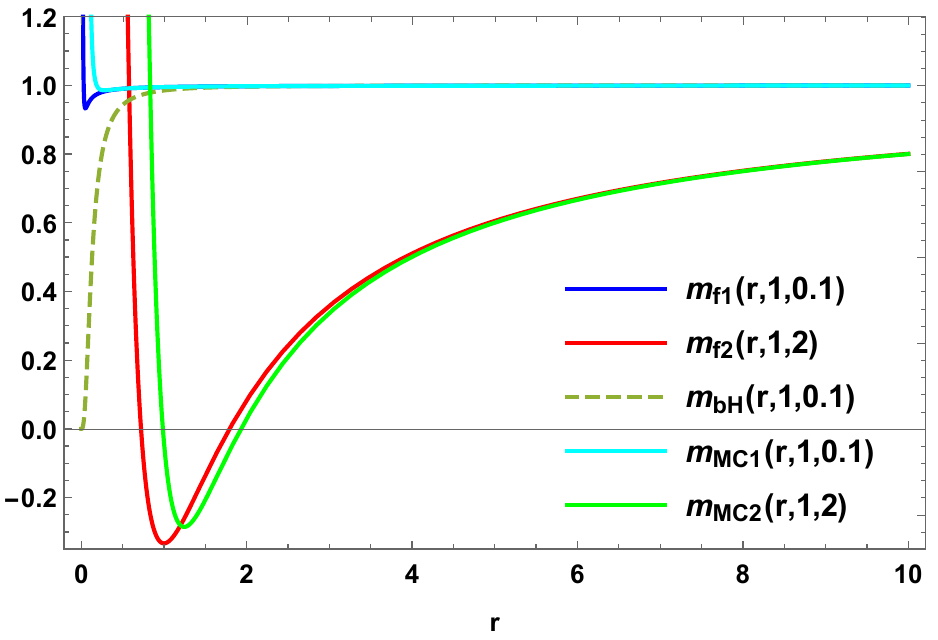}
\caption{Five mass functions $m_{f1}(r,m=1,q=0.1),~m_{bH}(r,1,0.1),m_{MC1}(r,1,0.1)$ are as functions of $r\in[0,10]$ and $m_{f2}(r,1,2),~m_{MC2}(r,1,2)$ as  functions of $r$. The first three are different inside the horizon, while they approach one $(m=1)$ outside the horizon. The last two have  negative region. }
\end{figure*}

This mass function will be  positive for $0<q<\frac{3m}{2}$, zero for $q=\frac{3m}{2}$, and negative for $q>\frac{3m}{2}$ (see Fig. 1).
In this case, the gauge field and derivative of the static scalar  take the forms
\begin{align}
A= \frac{q}{r}\Big(1-\frac{q^2}{6r^2}\Big)dt, \quad \phi'(r)=\sqrt{\frac{-q^2}{2\gamma r^2 f(r)}}
\end{align}
with $q$ an electric charge.
A negative coupling constant ($\gamma<0$) is chosen to have a real scalar hair  $\phi$. We note that  the scalar hair  is secondary  because it does not contain any independent scalar charge~\cite{Herdeiro:2015waa}.
It is worth noting  that the coupling constant $\gamma$ is included in the scalar hair only, but it does not appear in the metric functions.

From $g(r)=0$, one finds three analytic  solutions for the horizon
\begin{equation}
r_{1}(m,q),\quad r_{2}(m,q),\quad r_{i}(m,q)
\end{equation}
whose forms are too complicated to write down here. The remaining solution shows the negative horizon. The former is for describing the CHB(1), while the middle is for the NS(2).
In case of  $m=1$, one finds their explicit form as
\begin{eqnarray}
r_{1}(1,q)&=&  \frac{1}{2}\Bigg(1+\frac{\xi(q)}{\sqrt{3}}+\sqrt{2-\frac{4q^2}{3}-\frac{1}{3}q^{4/3}(-9+8q^2)^{1/3}+\frac{8\sqrt{3}(1-q^2)}{4\xi(q)}}\Bigg),    \label{h-11} \\
r_{2}(1,q)&=&   \frac{1}{2}\Bigg(1-\frac{\xi(q)}{\sqrt{3}}+\sqrt{2-\frac{4q^2}{3}-\frac{1}{3}q^{4/3}(-9+8q^2)^{1/3}-\frac{8\sqrt{3}(1-q^2)}{4\xi(q)}}\Bigg),   \label{h-22} \\
r_{i}(1,q)&=&   \frac{1}{2}\Bigg(1+\frac{\xi(q)}{\sqrt{3}}-\sqrt{2-\frac{4q^2}{3}-\frac{1}{3}q^{4/3}(-9+8q^2)^{1/3}+\frac{8\sqrt{3}(1-q^2)}{4\xi(q)}}\Bigg)    \label{h-33}
\end{eqnarray}
with
\begin{equation}
\xi(q)=\sqrt{3-2q^2+q^{4/3}(-9+8q^2)^{1/3}}.
\end{equation}
Here, we note that  $r_{i}(1,q)$ is a point but it shows up on  the photon sphere and shadow radius, implying the presence of its i-NS version.
Two singularities are found from $g(r)$ and $f(r)$: one is at $r=0$ and the other is at $r=r_{\rm NS}(q)=q/\sqrt{2}$, leading to the divergence of Kretschmann scalar defined by $R_{\mu\nu\rho\sigma}R^{\mu\nu\rho\sigma}$.
The weak cosmic censorship conjecture stating that the NS is behind the horizon implies the condition for mass and charge ($r_1(m,q)>q/\sqrt{2}$)
\begin{equation}
0<\frac{q}{m}<\sqrt{\frac{9}{8}}=1.06066(\simeq1.06),
\end{equation}
which defines $q_{\rm NS}=1.06m$.   For $q<q_{\rm NS}$, one finds the singular CHB(1), while for $q>q_{\rm NS}$, one has the NS(2) [see (Left) Fig. 2]. The $q=q_{\rm NS}$ corresponds to the singular (i) point.
This might imply that the inner horizon is replaced by the NS.

Now we are a position to introduce the  beyond Horndeski gravity (bHg)  which  respects both shift and parity symmetry  with $G=1$ as
\begin{eqnarray}
  \label{eq:act}
    {\cal L}_{\rm bHg}=\frac{1}{16\pi}\Big[G_4(X)R&+&G_{4X}\{(\square\phi)^2  -\phi_{;\mu\nu} \phi^{;\mu\nu}\}+G_2(X)\nonumber \\
  &+&F_4(X)\epsilon^{\mu\nu\rho\sigma}\epsilon^{\alpha\beta\gamma}_{~~~~\sigma}\phi_{;\mu}\phi_{;\alpha}\phi_{;\nu\beta}\phi_{;\rho\gamma}\Big].
\end{eqnarray}
Here, $G_4,~G_2,~F_4$ are arbitrary functions of   $X$,  the derivatives of the scalar field is defined as $\phi_{;\mu\nu}\equiv\nabla_\mu\partial_\nu\phi$, and a subscript $X$ denotes derivative with respect to $X$. The last term  of $F_4(X)\cdots$ represents beyond Horndeski gravity.
Concerning the scalar field, one  chooses
\begin{equation}
    \label{eq:phi}
    \phi(t,r)=\chi(t)+\psi(r), \quad \chi(t)=q t,
\end{equation}
where the linear time dependence is allowed because of the shift symmetry ($\phi\to \phi+$ const.) of Eq.(\ref{eq:act}).
A static and homogeneous  black hole solution was found with a primary scalar charge $\tilde{q}$  when choosing  $2XG_{4X}-G_4(X)+4X^2F_4(X)=-1$,
 $G_2=2b/\lambda^2 S(X)$, and $G_4(X)=1+b S(X)$ with $S(X)=c_{5/2}X^{5/2}$~\cite{Bakopoulos:2023sdm}. Here, $b$ and $\lambda$ represent  coupling constants and $X=(\tilde{q}\lambda)^2/(2r^2+2\lambda^2)$.
On the other hand, redefining $bc_{5/2}\to b$ and setting $b \lambda \tilde{q}^5 =-3\sqrt{2}M$ with $M$ mass of BH  leads to the beyond Horndeski BH (bH) solution  with its mass function $m_{bH}(r)$~\cite{Bakopoulos:2024ogt}
\begin{align} \label{R-bh}
    h(r)=1-\frac{2Mr^2}{(r^2+\lambda^2)^{3/2}}\equiv 1-\frac{2m_{bH}(r)}{r}.
\end{align}
Its secondary  scalar hair is given by
\begin{equation}
    \label{sec-h}
 \psi'(r)=\frac{\Big(\frac{3\sqrt{2}M}{b\lambda}\Big)^{1/5}}{h(r)}\sqrt{1-\frac{h(r)}{1+(r/\lambda)^2}},
\end{equation}
which is free from the scalar charge $\tilde{q}$.

At this stage,  we  note that Eq.(\ref{R-bh}) corresponds to the Bardeen  black hole without singularity at $r=0$ obtained from the Einstein-nonlinear electrodynamics (ENLED) theory whose Lagrangian is given by~\cite{Ayon-Beato:2000mjt}
 \begin{equation}\label{eq4:action}
 {\cal L}_{\rm ENLED}=\frac{1}{16 \pi}
  \Bigg[R  -\frac{6}{s \lambda^2}  \left(\frac{\sqrt{\lambda^2 \mathcal{F}/2}}{ 1 + \sqrt{\lambda^2 \mathcal{F}/2} }\right)^{\frac{5}{2}} \Bigg]
\end{equation}
with $s=|\lambda|/2M$. It seems that there is no direct connection between  Eq.(\ref{eq:act}) for bH and Eq.(\ref{eq4:action}). However, their effective energy-momentum tensors defined by $G_{\mu\nu}=T_{\mu\nu}$ are the same.
In the weak field limit, the NLED   takes the series  as
\begin{equation}
\frac{6 M \mathcal{F}^{5/4}}{2^{1/4}\lambda^{1/2}}\Big[1-\frac{5\lambda}{2\sqrt{2}} \mathcal{F}^{1/2}+\cdots\Big]_{\mathcal{F}\to2\lambda^2/r^4} =\frac{12M \lambda^2}{r^5}\Big[1-\frac{5\lambda^2}{2r^2}+\cdots\Big]\label{series-t}
\end{equation}
whose first term  is not   the Maxwell term of  $ \mathcal{F} $.
 In this case, $\lambda$ is the magnetic charge included as $F_{\theta\varphi}=\lambda\sin \theta$. Observing  Eqs.(\ref{eq4:action}) and (\ref{series-t}),
$M$ and $\lambda$  appear as coupling constants.
From now on, we replace $M$ and $\lambda$ by $m$ and $q$ for our purpose. As is shown in Fig. 1, the mass function $m_{bH}(r,m=1,q=0.1)$ is always zero at $r=0$ and $1(=m)$ near the event horizon at $r\simeq2$.
From six roots to $h(r)=0$, we obtain its event horizon as
\begin{equation}
r_{bH}(m,q)=\sqrt{\frac{1}{3}(4m^2-3q^2)+\frac{2^{1/3}}{3}\chi(m,q)+\frac{2^{2/3}(8m^4-12m^2q^2)}{3\chi(m,q)}}
\end{equation}
with
\begin{equation}
\chi(m,q)=\sqrt[3]{32m^6-72m^4\lambda^2+ 27m^2q^6+3\sqrt{81m^4q^8-48m^6q^6}}.
\end{equation}
It was shown that  two horizons (event/Cauchy horizons)  exist for $0<q/m<4/3\sqrt{3}=0.7698(\simeq 0.77)$  and two shrink into an extremal  BH at $q=0.77m$.
This implies that its allowed range is small as $q_{bH}\in[0,0.77]$ with $m=1$. So, it is necessary to introduce  another model to have a longer $q$-range.
\begin{figure*}[t!]
   \centering
   \includegraphics[width=0.4\textwidth]{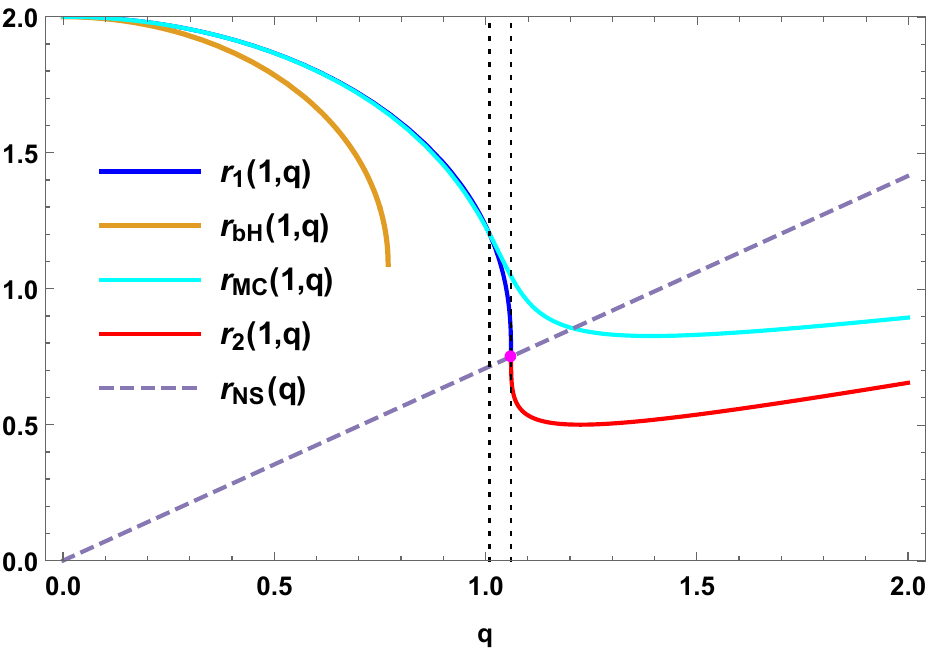}
   \hfill%
    \includegraphics[width=0.4\textwidth]{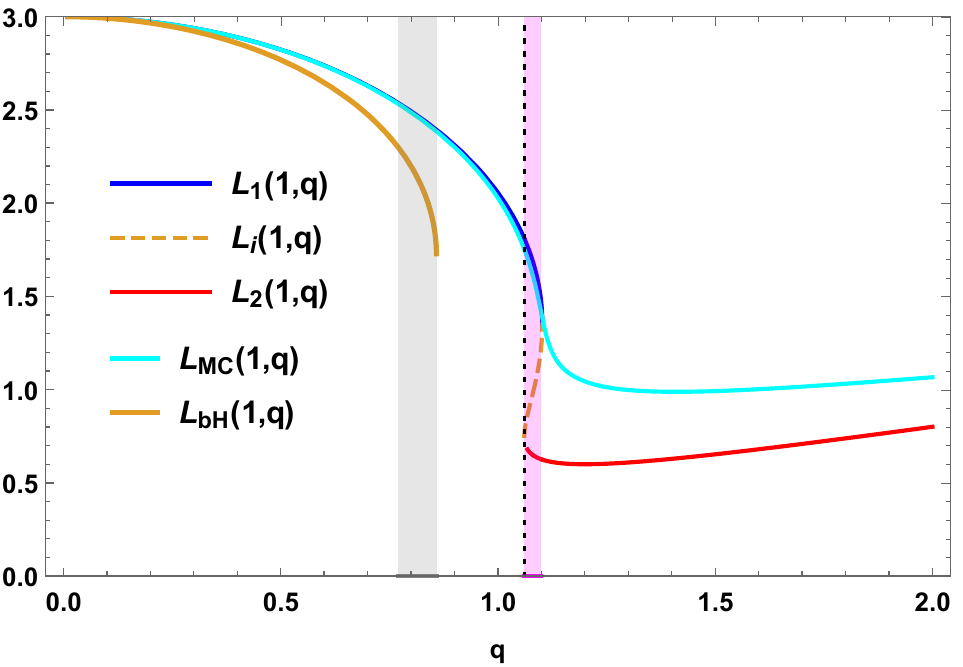}
\caption{ (Left) Four horizons  $r_{1}(1,q)$, $r_{bH}(1,q)$, $r_{MC}(1,q)$, and $r_{2}(1,q)$  as functions of $q$. Here, we introduce  a line of $r= r_{\rm NS}(q)$ for showing  the NS point. For $r> r_{NS}(q)$, one finds the CHB, whereas one has the NS for $r<r_{ NS}(q)$.  A magenta point at $(q=1.06,3/4)$ denotes $r_i(1,q)$. We note that $r_{MC}(1,q)$ is defined as  a single horizon without limitation of $q$. Two dotted lines are $r_1=r_{MC}$ at $q=1.01$ and $r_1=r_2$ at $q=1.06$. (Right) Five photon sphere radii  ($L_{1}(1,q),~L_{bH}(1,q)$) and  ($L_{2}(1,q),~L_{MC}(1,q)$)  with $L_i(1,q)$ as functions of $q$.  The first two are extended to include their NS versions defined in two shaded regions, compared to the last two. We note that $r_i(1,q)$ is realized as $L_{i}(1,q)$  being a connector between $L_1(1,q)$ and $L_2(1,q)$.  A dotted line denotes $q=1.06$ which is starting points for $L_2(1,q)$.   }
\end{figure*}

For this purpose, we introduce   a magnetically charged black hole (MC) with mass $m$ and magnetic charge $q$ obtained from an effective Lagrangian  for the Einstein-Euler-Heisenberg (EEH) theory~\cite{Yajima:2000kw}
\begin{equation}
{\cal L}_{\rm EEH}=\frac{1}{16\pi}\Big[R-(\mathcal{F}-\mu\mathcal{F}^2)\Big],\label{EEH-a}
\end{equation}
with a coupling constant $\mu=\frac{he^4}{360\pi^2m^4}$. It is interesting to note that the latter can be generated from the Born-Infeld action as~\cite{GarciaD:1984xrg}
\begin{equation}
\mathcal{L}_{BI}=4b^2\Big[-1+\sqrt{1+\mathcal{F}/2b^2}\Big]=\mathcal{F}-\frac{\mathcal{F}^2}{8b^2}+\cdots
\end{equation}
with $\mu=\frac{1}{8b^2}$. From the action (\ref{EEH-a}), one finds the single Einstein equation for the mass function
\begin{equation}
m'_{MC}(r)=\frac{q^2}{2r^2}-\mu \frac{q^4}{r^6} \label{massf-eq}
\end{equation}
with a gauge field $A_\varphi=-q \cos \theta$.
After integrating this equation,  the mass function and metric function take the forms
\begin{equation}
g_{\rm MC}(r)\equiv 1-\frac{2m_{MC}(r)}{r}=1-\frac{2m}{r}+\frac{q^2}{r^2}-\frac{2\mu}{5} \frac{q^4}{r^6},
\end{equation}
which is similar to $g(r)$ in Eq.(\ref{g-metricf}). Hereafter, we fix $\mu= 0.3$ for getting a singular BH with  a single horizon. In the case of $\mu\le 0.08$ with $m=1$, there exist four solution branches of the horizon~\cite{Myung:2025zxu}.  As is shown in Fig. 1, the mass function is similar to that for the CHB depending $q$.  From $g_{\rm MC}(r)=0$,  it has the  single horizon
\begin{equation}
r_{MC}(m,q)
\end{equation}
whose form is too complicated to show here.  From (Left) Fig. 2, one finds that $r_{MC}(1,q)$ is  a single horizon without limitation of $q$.  This is why we introduced this BH.  It is the nearly  same as $r_{1}(1,q)$ of the CHB until arriving the crossing point at $q_c=1.01$.  An interesting point  is
that there is no  theoretical constraint on restricting  the magnetic charge $q$ and thus, $r_{MC}(1,q)$ is a continuous function of $q$.  This point  differs from the separation of  CHB and NS at $q=1.06$ in the EHM theory.

Observing (Left) Fig. 2, there is no inner horizon for the CHB which states that it satisfies no scalar-haired inner horizon theorem~\cite{Devecioglu:2021xug} and a relevant branch is described by   $r_{1}(m,q_{1}\in[0,1.06))$.  Its $q$-range is larger than  $q_{RN}\in[0,1]$ of Reissner-Nordstr\"{o}m black hole (RN), which suggests that the CHB has a scalar hair $\phi(r)$. If the scalar hair exists, its scalarized  black hole is usually overcharged~\cite{Herdeiro:2018wub,Myung:2018vug}.
In particular, there exists the other branch described by $r_{2}(m,q_2\in(1.06,\infty))$~\cite{Myung:2025wmw}. However, it is always  under the  line $r_{\rm NS}(q)=q/\sqrt{2}$ and thus, it describes the NS only.
It is worth noting that  even though $r_{i}(1,q)$  is a magenta point at $q=1.06$  in (Left) Fig. 2,  its photon sphere  appears a connector between $r_1(1,q)$ and $r_2(1,q)$ [see (Right) Fig. 2],  implying the presence of its i-NS version.
For the regular bH, its $q$-range ($q_{bH}\in[0,0.77]$) is less than  $q_{RN}$ even though it possesses  a secondary scalar hair $\psi(r)$.  This is a curious point which one has to clarify.

\section{Shadow radius analysis}
To find the photon sphere of the CHB, we introduce the Lagrangian of the photon
\begin{equation}
{\cal L}_{\rm LP}=\frac{1}{2}g_{\mu\nu}\dot{x}^\mu\dot{x}^\nu=\frac{1}{2}\Big[-g(r)\dot{t}^2+\frac{\dot{r}^2}{f(r)}+r^2(\dot{\theta}^2+\sin^2\theta \dot{\varphi}^2)\Big].
\end{equation}
Taking the light traveling on the equational plane of the  CHB ($\theta=\pi/2$ and $\dot{\theta}=0$) and considering a spherically symmetric and static metric Eq.(\ref{metric-ansatz}),
there exist two conserved quantities of photon (energy and angular momentum) as
\begin{equation}
E=-\frac{\partial {\cal L}_{LP}}{\partial \dot{t}}=g(r)\dot{t},\quad \tilde{L}=\frac{\partial {\cal L}_{LP}}{\partial \dot{\varphi}}=r^2\dot{\varphi}.
\end{equation}
Taking into account the null geodesic for the photon ($ds^2=0$) with the affine parameter $\tilde{\lambda}=\lambda \tilde{L}$ and impact parameter $b=\tilde{L}/E$, its radial equation of motion is  given by
\begin{equation}
\frac{dr}{d\tilde{\lambda}}=\sqrt{\frac{f(r)}{b^2 g(r)}-\frac{f(r)}{r^2}}.
\end{equation}
In this case, the effective potential for a photon takes the form
\begin{equation}
V(r)=\frac{1}{2b^2}-\frac{f(r)}{2}\Big[\frac{1}{b^2 g(r)}-\frac{1}{r^2}\Big].
\end{equation}
Requiring the photon sphere ($\dot{r}=0,~\ddot{r}=0$), one finds two conditions
\begin{equation} \label{cond-LR}
V(r=L)=\frac{1}{2b_L^2}, \quad V'(r=L)=0,
\end{equation}
where $b_L$ denotes the critical impact parameter and $L$ represents the radius  of photon sphere.
\begin{figure*}[t!]
   \centering
  \includegraphics[width=0.4\textwidth]{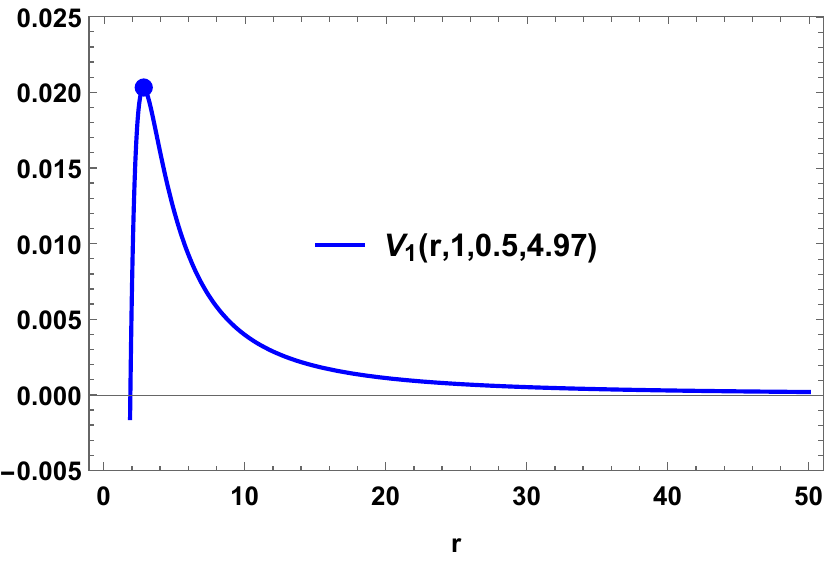}
 \hfill%
    \includegraphics[width=0.4\textwidth]{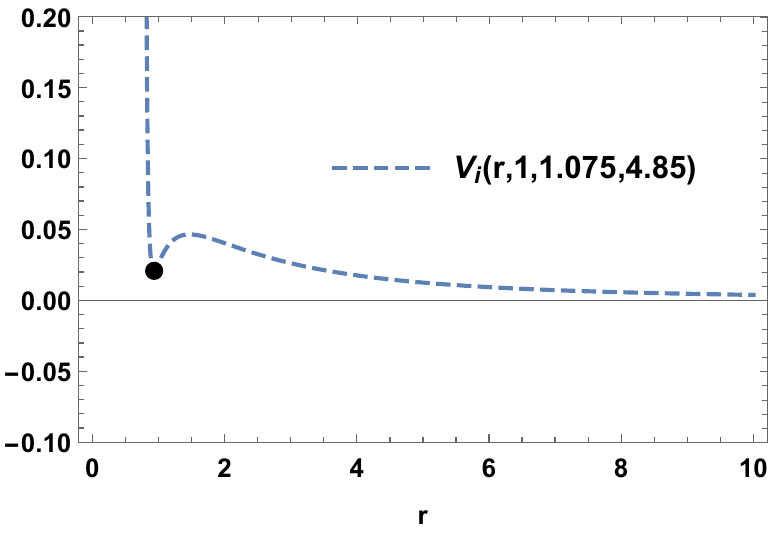}
\caption{ Two potentials $V_1(r,m=1,q=0.5,b_1=4.97)$ and $V_i(r,1,1.075,b_i=4.85)$ as functions of $r\ge r_{1,i}(1,q)$. One point (blue) corresponds to an unstable point at $r=L_1(1,0.5)=2.82$ and the other (black) denotes a stable point at $r=L_i(1,1.075)=0.93$.
The right one has a positive blow-up point at $r=\frac{3}{4}$ [a magenta point in (Left) Fig. 2] and an unstable point at  $r=1.47$.  }
\end{figure*}
\begin{figure*}[t!]
   \centering
  \includegraphics[width=0.3\textwidth]{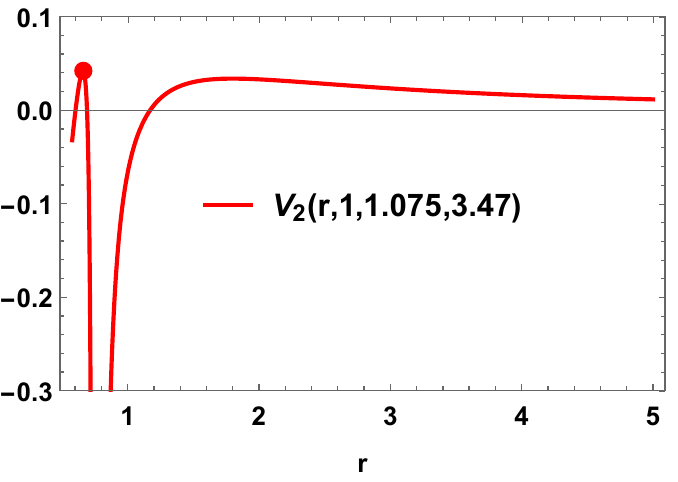}
 \hfill%
    \includegraphics[width=0.3\textwidth]{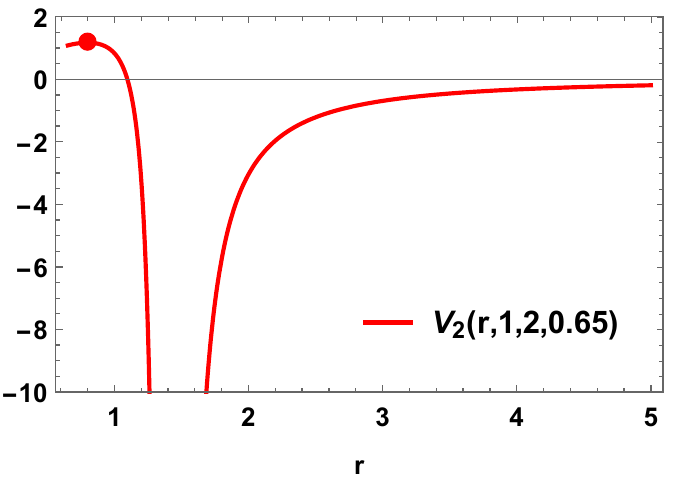}
    \hfill%
    \includegraphics[width=0.3\textwidth]{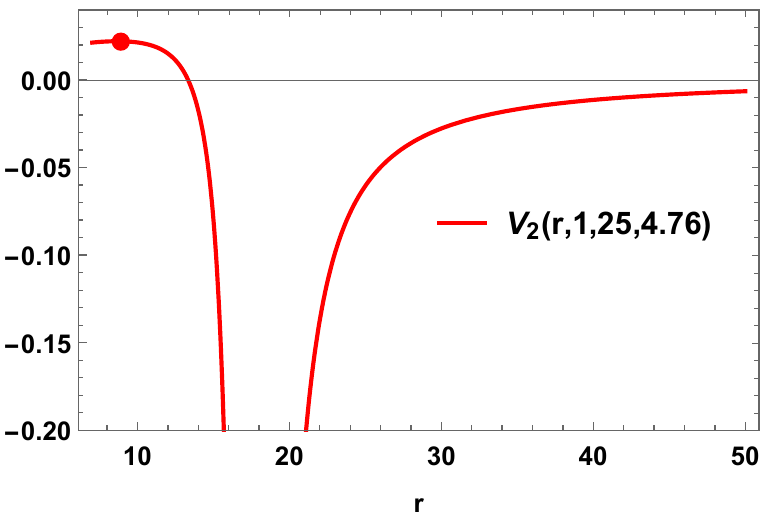}
\caption{ Three potentials $V_2(r,m=1,q=1.075,b_2=3.47),~V_2(r,1,2,b_2=0.65)$ and $V_2(r,1,25,b_2=4.76)$ as functions of $r\ge r_2(1,q)$. Three points correspond to unstable points at $r=L_2(1,q)=0.66,0.8,8.87$. The left one has an additional unstable point at $r=1.8$.  Also, these  all have  (negative) blow-up points at $r=0.71 q$, indicating  a feature of $V_2$. }
\end{figure*}
As is shown in Fig. 3, we display one unstable and  one stable points in the effective potentials with  $q=0.5,~1.075$. Fig. 4 indicates three unstable points  in the effective potentials with  $q=1.075,~2,~25$.
The (Left) Fig. 3 shows a conventional potential but the (Right) Fig. 3 indicates a peculiar potential with a blow-up point for realizing from a magenta point in (Left) Fig. 2. Also, Fig. 4 shows peculiar potentials with (negative) blow-up point to represent their NS nature. These characteristic behaviors predict the peculiar forms for the critical impact parameters.

 Eq.(\ref{cond-LR}) implies two relations
 \begin{equation}
 L^2=g(L)b_L^2,\quad 2g(L)=Lg'(L).
 \end{equation}
Here, we find three photon spheres and their critical impact parameters for the CHB(1),  NS(2), and i-NS(i) as
\begin{eqnarray}
&&L_{1}(m,q),\quad L_2(m,q),\quad  L_{i}(m,q),\label{LR} \\
&&b_{1}(m,q),\quad b_{2}(m,q),\quad b_i(m,q),\label{CI}
\end{eqnarray}
whose explicit forms are too complicated to write down here.

Furthermore, two photon spheres and their critical impact parameters for the singular MC  and regular bH are given by
\begin{eqnarray}
&&L_{MC}(m,q),\quad L_{bH}(m,q), \label{LR} \\
&&b_{MC}(m,q),\quad b_{bH}(m,q),\label{CI}
\end{eqnarray}
where $L_{MC}(m,q)$  and $L_{bH}(m,q)$ are   obtained from solving $ 2g_{MC}(L_{MC})=L_{MC}g'_{MC}(L_{MC})$ and $2h(L_{bH})=L_{bH}h'(L_{bH})$. $b_{MC}(m,q)$ and  $b_{bH}(m,q)$ are found from
\begin{equation}
b_{MC}=\frac{L_{MC}}{\sqrt{g_{MC}(L_{MC})}},\quad b_{bH}=\frac{L_{bH}}{\sqrt{h(L_{bH})}}.
\end{equation}
\begin{figure*}[t!]
   \centering
  \includegraphics[width=0.5\textwidth]{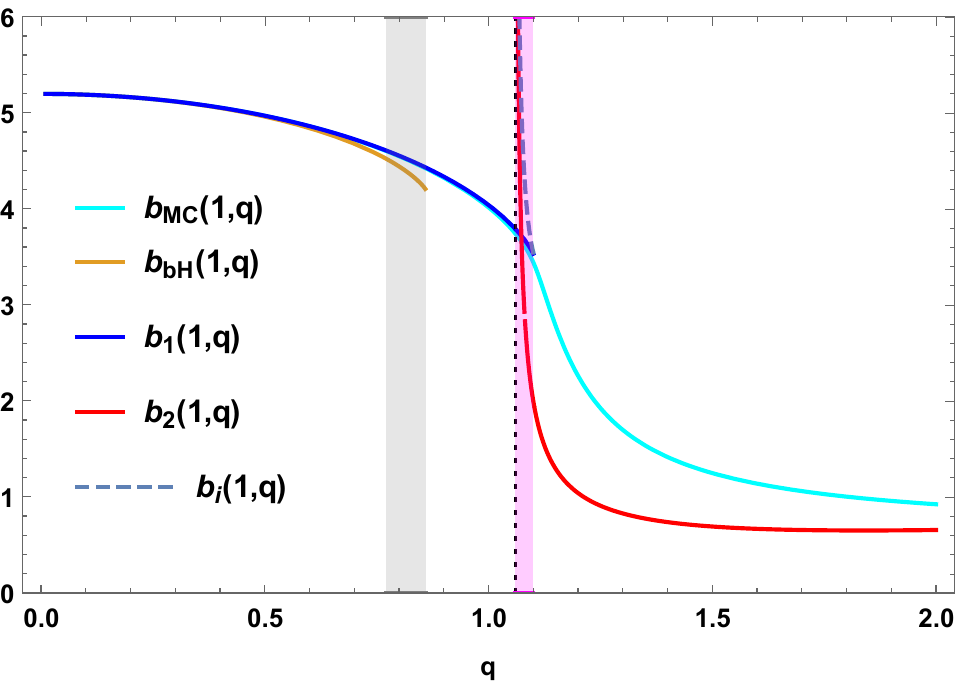}
\caption{ Five shadow radii $ b_{MC}(1,q)$,  $b_{bH}(1,q)$,  $b_{1}(1,q)$,  $ b_{2}(1,q)$, and  $b_{i}(1,q)$ as functions of $q\in[0,2]$.  It is found that $b_{MC}(1,q) \simeq b_1(1,q)$ for $q\in[0,1.1]$. A dotted line denotes $q=1.06$ which is a blow-up point for $b_{2}(1,q)$ and $b_{i}(1,q)$.  Here,  $q_{bH-NS}\in(0.77,0.86]$ (gray column) accommodates  the bH-NS and $q\in(1.06,1.1]$ (magenta column) includes  the CHB-NS and i-NS.  }
\end{figure*}

(Right) Fig. 2 shows the photon sphere radii and Fig. 5 represents shadow radii.
From $L_1(1,q)$ and $b_1(1,q)$, we find that its allowed $q$-range is extended from $[0,\frac{3}{2\sqrt{2}}=1.06]$ to $[0,1.1002\simeq 1.1]$. Here, the upper limit of 1.10 is determined by the existence condition of $L_1(1,q)$.
This implies that $q_{1-NS}\in(1.06,1.10]$ (magenta column) denotes the CHB-NS. We note that $L_{i}(1,q)$ is present as a NS-connector appearing  between $L_1(1,q)$ and $L_2(1,q)$.
But, there is no change for the NS: $q_{2}\in(1.06,\infty)$ and the MC: $q_{MC}\in[0,\infty)$.
Also, from analyzing  $L_{bH}(1,q)$ and $b_{bH}(1,q)$, one finds  that  its $q$-range is extended  from $[0,\frac{4}{3\sqrt{3}}=0.77]$ to $[0,\frac{48}{25\sqrt{5}}=0.86]$. This means  that  $q_{bH-NS}\in(0.77,0.86]$ (gray column) denotes the bH-NS. Importantly, the dotted line in Fig. 5 indicates  the  blow-up  point for $b_2(1,q)$ and $b_i(1,q)$, conjecturing from their effective potentials in Figs. 3 and 4. However, $b_1(1,q)$ crosses this line to get its NS version (CHB-NS).

\section{Test with EHT observation}

From the EHT observation (Keck- and VLTI-based estimates for SgrA$*$~\cite{EventHorizonTelescope:2022wkp,EventHorizonTelescope:2022wok,EventHorizonTelescope:2022xqj}), the  $1\sigma$ constraint on the shadow radius $r_{\rm sh}=b_L$ is given by~\cite{Vagnozzi:2022moj}
\begin{equation}
4.55\lesssim r_{\rm sh} \lesssim 5.22  \label{KV1}
\end{equation}
and the  $2\sigma$ constraint implies
\begin{equation}
4.21 \lesssim r_{\rm sh} \lesssim 5.56. \label{KV2}
\end{equation}
Fig. 6 indicates for explicit graphes to compare with the EHT observation.
For the CHB ($q_{1}\in [0,1.06]$), one has two constraints of the upper limits on its electric charge $q$: $q\lesssim 0.801 (1\sigma)$ including a blue dot and $0.946 (2\sigma)$.  However,  the CHB-NS ($q_{1-NS}\in(1.06,1.10]$: magenta column) including a red dot is completely ruled out from its $2\sigma$.
The EHT observation rules out the possibility of SgrA$*$ being the singular and extremal point of CHB ($q_{NS}=1.06$).
For the MC ($q_{MC}\in[0,\infty$)),  we have   two constraints on its magnetic charge:  $q\lesssim 0.799 (1\sigma)$ and $0.941 (2\sigma)$ which  are the nearly same as for the CHB.  There is no extremal MC from the EEH theory.
Also, it is meaningful to note that there is no $q>1$ branch which constrains its magnetic  charge because its shadow radius is a monotonically decreasing function of  $q$.
For the i-NS branch whose horizon is a magenta point in (Left) Fig. 2, we find two narrow constraints: $1.072\lesssim q \lesssim 1.078 (1\sigma)$ including a black dot and $1.071\lesssim q \lesssim 1.082(2\sigma)$ on  the electric charge $q$. Here, we include a black dot at $q=1.075$ located within $1\sigma$.
\begin{figure*}[t!]
   \centering
  \includegraphics[width=0.4\textwidth]{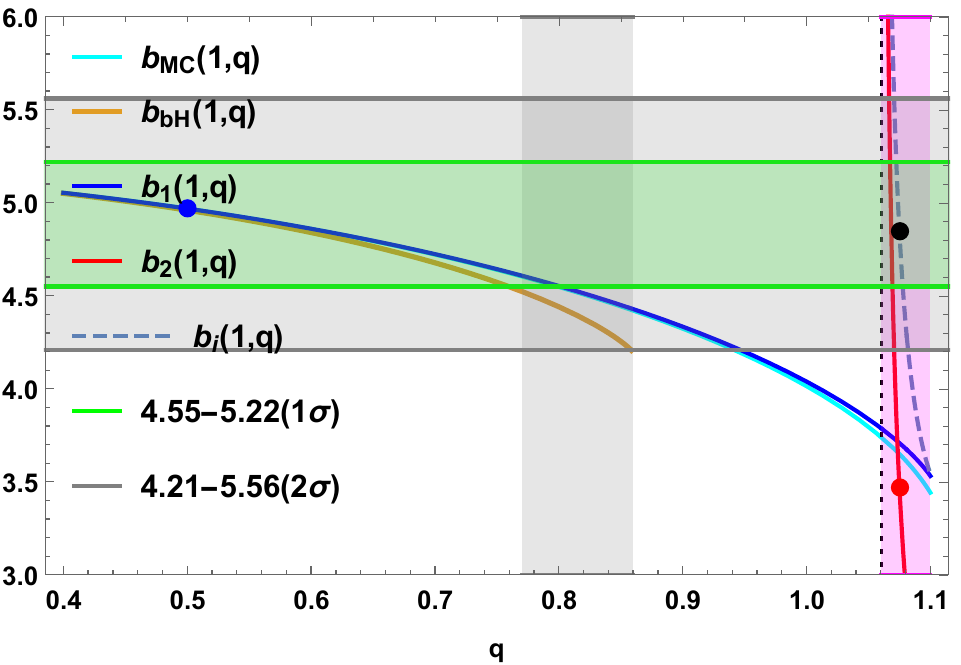}
 \hfill%
    \includegraphics[width=0.4\textwidth]{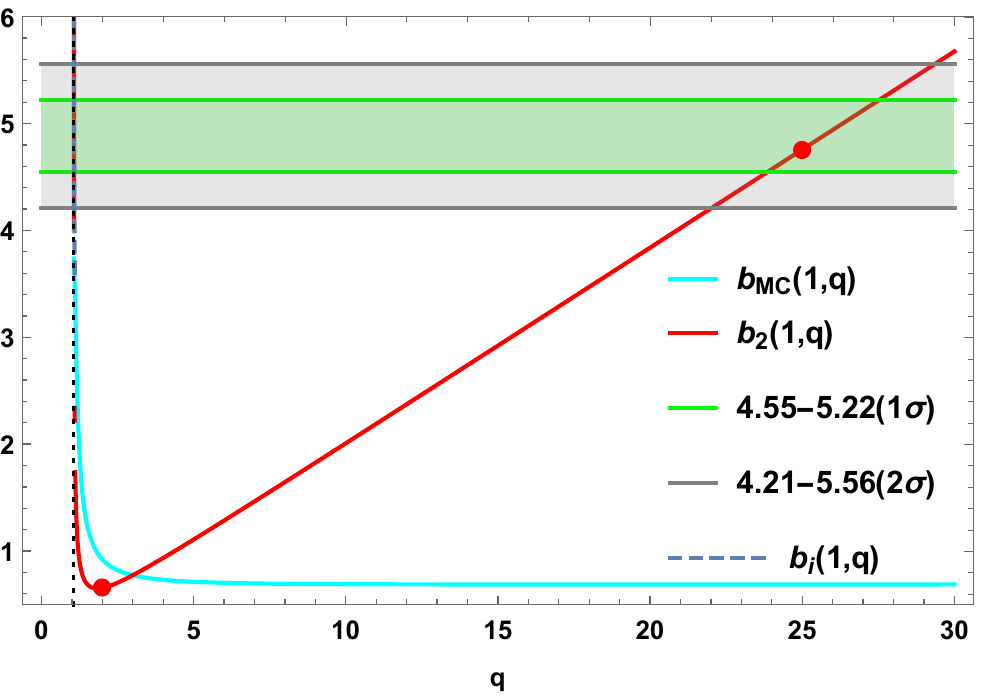}
\caption{(Left).  Five shadow radii  $ b_{MC}(1,q), b_{bH}(1,q)$, $ b_{1}(1,q)$, $b_2(1,q)$, and $b_i(1,q)$ as functions of $q\in[0,1.2]$.  Here, we introduce $1\sigma$ and $2\sigma$ ranges. Three dots  are  the blue  and black dots appeared in Fig. 3, and a red dot appeared in (Left) Fig. 4.  (Right)   $b_2(1,q)$ is a function of $q\in(1.06,30]$ for the NS and it has two branches. $b_{MC}(1,q)$ is a monotonically decreasing function of $q$.  Two dots indicate the red points  appeared in (Middle and Right) Fig. 4.  The dotted line ($q=1.06$) denotes the blow-up point for $b_2(1,q)$ and $b_i(1,q)$ while $b_{MC}(1,1.06)=3.74$.   }
\end{figure*}

On the other hand, the regular bH ($q_{bH}\in[0,0.77]$) is constrained as $q\lesssim 0.76 (1\sigma)$ for its coupling constant $q$ but it is unconstrained from $q\lesssim 0.86(2\sigma)$.
The bH-NS ($q_{bH-NS}\in(0.77,0.86]$: gray column) is unconstrained by $q\lesssim 0.86(2\sigma)$.  This implies that within $2\sigma$, the bH and bH-NS including an extremal bH ($q_{ebH}=0.77$) are consistent with the EHT observation~\cite{Vagnozzi:2022moj}.
Finally,  the NS ($q_{NS}\in(1.06,\infty)$) have two branches. There is no constraint from  one branch   existing on the magenta column  because it is a nearly vertical line.
 The other of a linearly increasing branch  is  constrained as $23.8 \lesssim q \lesssim 27.5 (1\sigma)$ including a red dot and $22\lesssim q \lesssim 29 (2\sigma)$ for its electric charge $q$, which is intriguing but seems extreme. This corresponds to a new feature of the NS found from the EHM theory.  However, this might not be physically plausible but imply a limitation of the NS with unlimited charge $q$ found from the EHM theory.

\section{Classical scattering analysis}

We know that the critical impact parameters  $b_2(1,q)$ and $b_i(1,q)$ take  peculiar forms, compared with others.
These came from the i-NS and NS, differing from BHs. We note that $b_{MC}(1,q)$ is a continuously decreasing function of $q$.  To understand them, we need to introduce the scattering picture.
Scattering of a scalar field off a BH is an interesting topic~\cite{Benone:2017hll}.
Scattering and absorption of scalar ﬁelds by various BHs were already studied for Schwarzschild BH~\cite{Sanchez:1977si}, RN~\cite{Crispino:2009ki}, Kerr BH~\cite{Glampedakis:2001cx},  regular BH~\cite{Macedo:2014uga}, and singular Einstein-Euler-Heisenberg BH~\cite{Olvera:2019unw}.
In the classical (high-frequency) limit, absorption (geometric) cross section of a scalar field  is directly represented by the critical impact parameter as
\begin{equation}
\sigma_{cL}(m,q)=\pi b^2_L(m,q),
\end{equation}
which means that  we may infer their property  by considering the geodesic scattering. An improvement of the high-frequency cross section was proposed with the  oscillatory part in the eikonal limit~\cite{Decanini:2011xi}.

There was a recently scattering study  focused on the CHB(1) by considering a charged scalar propagation~\cite{Li:2024xyu}, but there is no analysis on the i-NS and NS.
\begin{figure*}[t!]
   \centering
  \includegraphics[width=0.4\textwidth]{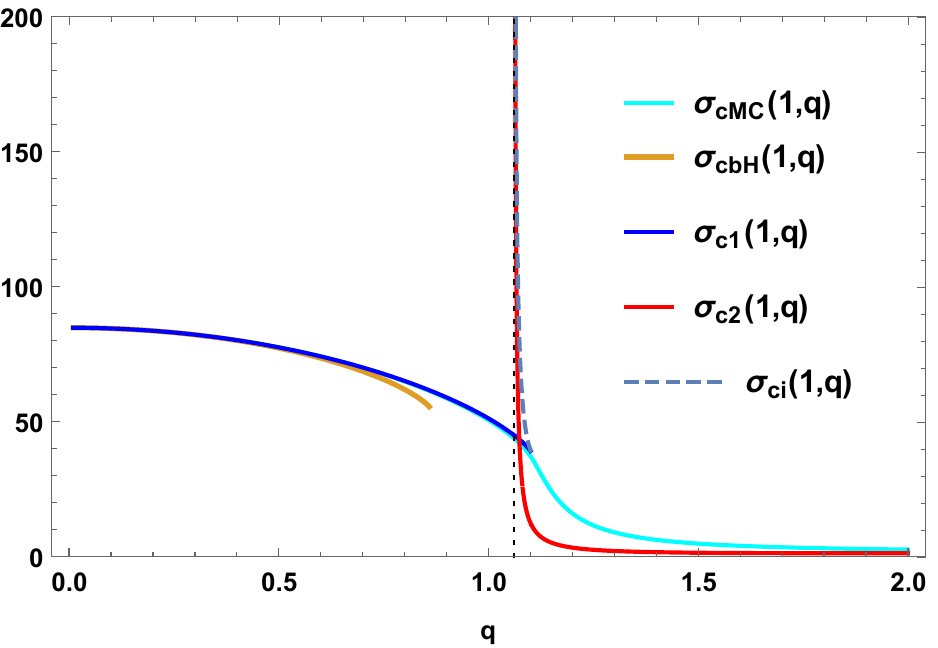}
 \hfill%
    \includegraphics[width=0.4\textwidth]{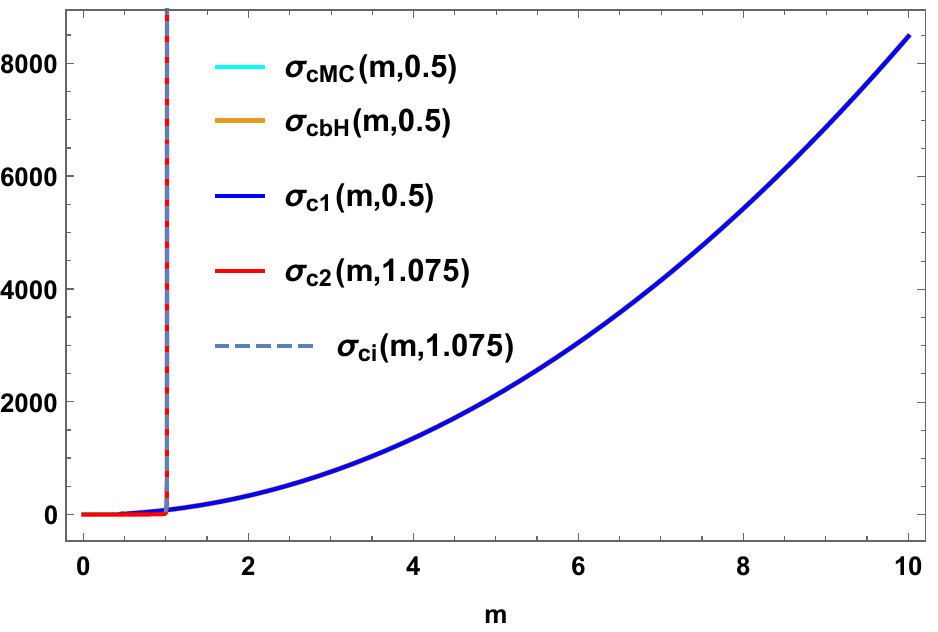}
\caption{(Left).  Five geometric cross sections  $ \sigma_{cMC}(m=1,q), \sigma_{cbH}(1,q)$, $ \sigma_{c1}(1,q)$, $\sigma_{c2}(1,q)$, and $\sigma_{ci}(1,q)$ as functions of charge $q\in[0,2]$. The dotted line is a blow-up point at $q=1.06$. (Right)   Five geometric cross sections  $ \sigma_{cMC}(m,0.5), \sigma_{cbH}(m,0.5)$, $ \sigma_{c1}(m,0.5)$, $\sigma_{c2}(m,1.075)$, and $\sigma_{ci}(m,1.075)$ as functions of mass $m\in[0,10]$. The last two are divergent at $m=1.01$.   }
\end{figure*}
As was shown in Fig. 7, one finds that  $ \sigma_{cMC}(m=1,q) \simeq \sigma_{cbH}(1,q)\simeq  \sigma_{c1}(1,q)$, implying decreasing functions of  $q$~\cite{Macedo:2014uga} while   $ \sigma_{cMC}(m,q=0.5) \simeq \sigma_{cbH}(m,0.5)\simeq  \sigma_{c1}(m,0.5)$ for $m\in [0,10]$, indicating increasing functions of  $m\in [0,10]$. This  shows  a promising behavior of  geometric cross sections for BH with/without singularity.
However, one finds that $\sigma_{c2}(1,q)$ and $\sigma_{ci}(1,q)$ blow up at $q=1.06$, and $\sigma_{c2}(m,q=1.075)$ and $\sigma_{ci}(m,q=1.075)$ are divergent at $m=1.01$.
This provides us a hint of scattering behavior when waves (lights) are scattered off the i-NS and NS.  In the case of $b>b_L$ for MC, bH, and CHB(1), the particles scatter off the center and the gravitational captures do not happen. For the above cases of $q=1.06$ and $m=1.01$, however,  all particles pass into the i-NS and NS and thus, they all are captured by i-NS and NS.
Although the contrast between BH and NS scattering behaviors is compelling,  it would be better to consider  observational features of lensing or accretion disk  to understand  the distinction BH and NS.

Finally, we wish to mention two limiting cases of $q$.  One is    $\sigma_{cMC}(1,q\to\infty)=0$ which means that all particles are scattered off the center (point) and the gravitational captures never happen. The other is  $\sigma_{c2}(1,q\to\infty)=\infty$ which implies that all particles are absorbed by the center  and the gravitational captures always happen.

\section{Discussions}

The shadow radii of  various BH and NS found from modified gravity theories  were extensively  used to test the EHT results for SgrA$^*$ BH~\cite{EventHorizonTelescope:2022wkp,EventHorizonTelescope:2022wok,EventHorizonTelescope:2022xqj} and thus, to constrain their hair parameters~\cite{Vagnozzi:2022moj,daSilva:2023jxa}.
Recently, the CHB was employed to investigating shadow images with  distinctive  thin accretions~\cite{Gao:2023mjb}, weak and strong gravitational lensings~\cite{Wang:2019cuf}, and scattering cross section~\cite{Li:2024xyu}.

In this work, we test the CHB, CHB-NS, MC, i-NS, bH, bH-NS, and the NS with the EHT observation for SgrA$^*$ by computing their shadow radii (critical impact parameters).
For the CHB from the EHM theory, one has two constraints of the $1\sigma$ upper limit  $q\lesssim 0.801m$ and $2\sigma$ upper limit $q\lesssim 0.946m$.  However,  its NS version (CHB-NS: magenta column) is completely ruled out from its $2\sigma$.
The EHT observation rules out the possibility of SgrA$*$ being the singular and extremal point of CHB ($q_{NS}=1.06$).
For the MC ($q_{MC}\in[0,\infty$)) from the EEH theory,  we have   two constraints of $q\lesssim 0.799 (1\sigma)$ and $0.941 (2\sigma)$, which are the nearly same as in the CHB. There is no $q>1$ branch (NS version) which constrains its charge because $b_{MC}(1,q)$ is a monotonically decreasing function of $q$.
Within $2\sigma$, the bH without singularity and bH-NS (gray column) including an extremal bH ($q_{ebH}=0.77$) are consistent with the EHT observation~\cite{Vagnozzi:2022moj}.

On the other hand,  the NS obtained from the EHM theory is  constrained as $23.8 \lesssim q \lesssim 27.5 (1\sigma)$ and $22\lesssim q \lesssim 29 (2\sigma)$, showing a new feature of the NS.
For the i-NS branch whose horizon is a point, we found two narrow constraints: $1.072\lesssim q \lesssim 1.078 (1\sigma)$ and $1.071\lesssim q \lesssim 1.082(2\sigma)$ on  the electric charge $q$.

From classical scattering analysis, one found  that CHB, MC, and  bH with/without singularity at $r=0$ are quite different from  i-NS and NS.

\vspace{1cm}

{\bf Acknowledgments}
 \vspace{1cm}

 This work was supported by the National Research Foundation of Korea (NRF) grant
 funded by the Korea government(MSIT) (RS-2022-NR069013).

\end{document}